\documentclass[12pt,preprint]{aastex}

\slugcomment{Submitted, \today}

\shorttitle{MASS/DIMM Seeing at PEARL}
\shortauthors{Steinbring et al.}

\input epsf
\def\plotone#1{\centering \leavevmode
\epsfxsize=1.0\columnwidth \epsfbox{#1}}
\def\plotonehalf#1{\centering \leavevmode
\epsfxsize=0.5\columnwidth \epsfbox{#1}}
\def\plotonenarrow#1{\centering \leavevmode
\epsfxsize=0.75\columnwidth \epsfbox{#1}}

\DeclareMathAlphabet{\mathscr}{OT1}{pzc}{m}{it}

\begin{document}

\title{Preliminary DIMM and MASS Nighttime Seeing Measurements at PEARL, in the Canadian High Arctic}

\author{Eric Steinbring\altaffilmark{1}, Max Millar-Blanchaer\altaffilmark{2}, Wayne Ngan\altaffilmark{2}, Rick Murowinski\altaffilmark{1}, Brian Leckie\altaffilmark{1} \& Ray Carlberg\altaffilmark{2}}

\altaffiltext{1}{National Science Infrastructure, National Research Council Canada, Victoria, BC V9E 2E7, Canada}

\altaffiltext{2}{Department of Astronomy and Astrophysics, University of Toronto, Toronto, ON M5S 3H4, Canada}

\begin{abstract}

Results of deploying a Differential Image Motion Monitor (DIMM) and a DIMM combined with a Multi-Aperture Scintillation Sensor (MASS/DIMM) are reported for campaigns in 2011 and 2012 on the roof of the Polar Environment Atmospheric Research Laboratory (PEARL). This facility is on a 610-m-high ridge at latitude 80\arcdeg N, near the Eureka weatherstation on Ellesmere Island, Canada. The median seeing at 8-m elevation is 0\farcs85 or better based on DIMM data alone, but is dependent on wind direction, and likely includes a component due to the PEARL building itself. Results with MASS/DIMM yield a median seeing less than 0\farcs76. A semi-empirical model of seeing versus ground wind speed is introduced which allows agreement between these datasets, and with previous boundary-layer profiling by lunar scintillometry from the same location. This further suggests that best 20\%-ile seeing reaches 0\farcs53, of which typically 0\farcs30 is due to the free atmosphere. Some discussion for guiding future seeing instrumentation and characterization at this site is provided.

\end{abstract}

\keywords{site testing; instrumentation}

\section{Introduction}\label{introduction}

\begin{figure}
\plotone{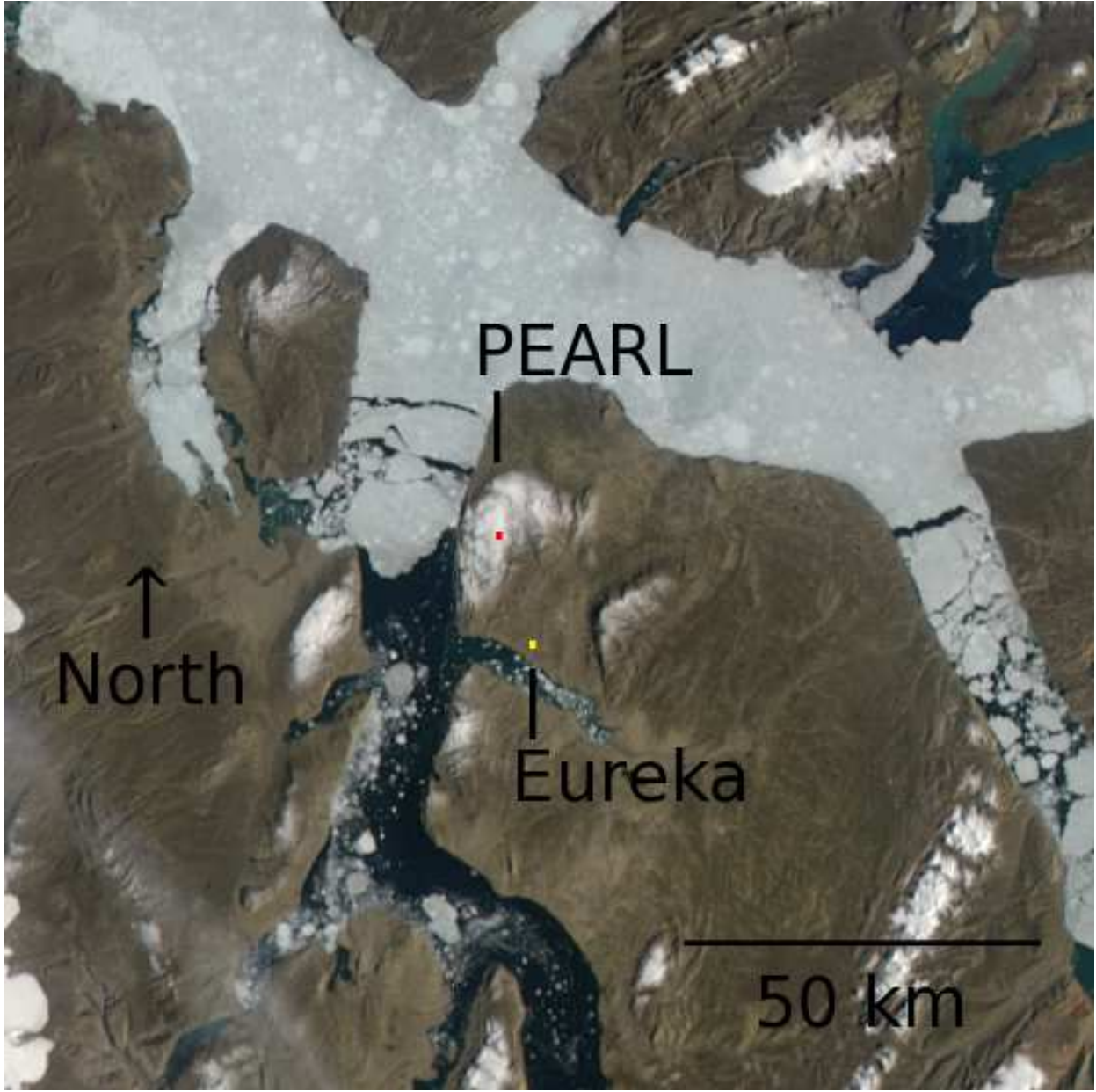}
\caption{A satellite image of the Fosheim Peninsula on Ellesmere Island, Canada, showing Eureka (indicated by a dot; yellow in the online version) and PEARL (red). The Arctic Ocean is to the northwest, at the mouth of Nansen Sound, upper left.}
\label{figure_fosheim_peninsula}
\end{figure}

Recent site testing of mountains on Ellesmere Island in the Canadian High Arctic indicates that they possess good astronomical sky qualities \citep{Steinbring2010}. The main reasons for this are their cold and dry environment, and a strong, persistent atmospheric thermal inversion peaked at $\sim 1000$ m elevation that develops during darkness at these latitudes. Thus, it is possible for relatively low-lying terrain to rise above the worst deleterious effects for optical/infrared astronomy: thick clouds and turbulent air. High clear-sky fractions have been shown for the Polar Environment Atmospheric Research Laboratory (PEARL; 80\arcdeg N, see Figure~\ref{figure_fosheim_peninsula}) atop a 610 m ridge near the Eureka weather station, which usually enjoys mild winds in winter \citep{Steinbring2012a}. A further expectation is that the PEARL site and other isolated terrain up to 1900 m elevation along the nearby ice-locked coast may also provide excellent seeing \citep{Steinbring2012b}. Some justification for that view comes from analogy to the Antarctic glacial plateau, where the free-atmospheric (FA) seeing is known to be exceptional \citep{Lawrence2004}. The case is plausibly the same in the High Arctic, as above the thermal inversion the temperature and wind profiles near the Poles are comparable. This neglects boundary-layer seeing, which for South Pole and Dome C is poor near the flat ice surface of the plateau, typically over 1\arcsec~at 8 m height \citep{Lawrence2010, Aristidi2013}. Conversely, lunar scintillometry obtained at PEARL with the Arctic Turbulence Profiler \citep[ATP;][]{Hickson2010, Hickson2013} indicates that it is within a thin, weak boundary layer: median under 0\farcs72 at 7 m, more akin to best midlatitude sites.

The PEARL facility, although not intended for astronomy, was designed for atmospheric physics measurements with optical instruments. It has a $7~{\rm m}\times18~{\rm m}$ observing platform over the central part of its 6-m high flat roof, unobstructed on all sides apart from a safety railing and a $3~{\rm m}\times3~{\rm m}$, 2.5-m tall structure at the northeast corner - an instrument enclosure referred to as the ``penthouse."  A sonic anemometer on a mast at the center of the roof reports barometric pressure, temperature, wind direction and speed at 10 m above ground. The building provides warm work space, power, and broadband satellite communications; roof access is by an outside stairwell. The arrangement allows operation of various site-testing instruments during campaigns. This paper presents first efforts to deploy two Differential Image Motion Monitor (DIMM) units, including one combined with Multi-Aperture Scintilllation Sensor (MASS/DIMM). The intention was to gain some insight into integrated seeing conditions, allow comparison with ATP, and to guide development of an autonomous MASS/DIMM seeing monitor to be deployed nearby.

Both DIMM and MASS are commonly used for site testing, with well-understood measurement properties. The DIMM \citep[][for a recent review]{Sarazin1990, Tokovinin2002} uses a subaperture mask and prisms to create multiple images of a single star. The relative displacements of these spots in the focal plane are measured via short-exposure images with a CCD, and converted to an estimate of seeing integrated through the full atmosphere for a large-aperture telescope. It is complementary to the MASS \citep[see][]{Tokovinin2007}, which is not sensitive to local seeing. The MASS apodizes the telescope pupil, separating the light of a star into four concentric rings, and measures its scintillation at millisecond cadence using photomultiplier tubes. This allows a profile of $C_n^2$ in 6 broad bins from 500 m to 16 km. A measure of 0\farcs27 FA seeing originally reported above Dome C was detected with MASS \citep{Lawrence2004}. The instruments and their implementation at PEARL are described in Section~\ref{instrumentation}; details of data collection and reductions are discussed in Section~\ref{observations}. An analysis of the results follows in Section~\ref{analysis}, followed by a summary and conclusions in Section~\ref{summary}.

\section{Instruments and Deployment}\label{instrumentation}

One DIMM was deployed at PEARL in 2011 February. This was followed in 2012 February with the addition of MASS/DIMM, which was re-deployed the following October. By mid-October, strong thermal-inversion conditions prevail above Eureka until the end of darkness, beginning in March. To ease intercomparison of results, the instruments were set up as much as possible to be the same as at other sites, with some special considerations due to the location. These will be highlighted below, along with the potential influence of environment and weather on the sampling.

\subsection{Optical and Electronic Setup}

\begin{figure}
\plotone{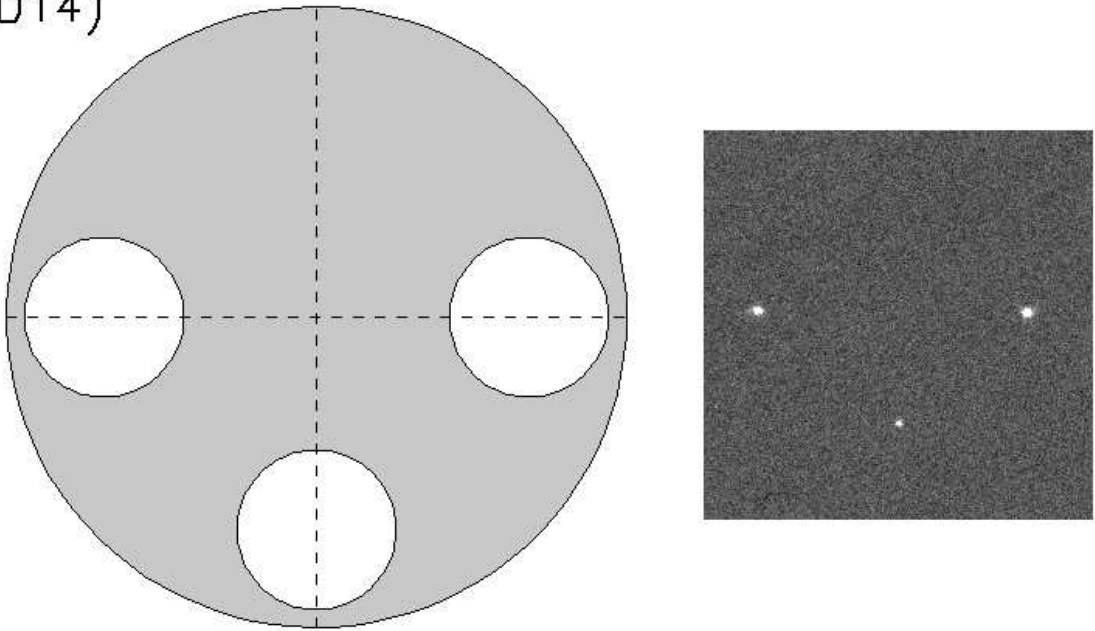}
\caption{A diagram of the triangular sub-pupil mask geometry for D14; one open aperture (at bottom) and two with prisms, all of 90 mm diameter and equidistant from the optical axis. The separation between the center of the open aperture and each with a prism is 170 mm; resultant $200\times200$ pixel frame at the focal plane is shown at right.}
\label{figure_d14_geometry}
\end{figure}

The original ``14-inch DIMM" (hereafter abbreviated D14) first deployed in 2011 used a portable 35 cm aperture Meade LX200 3556 mm focal-length Schmidt-Cassegrain telescope and a Moravian Instruments G1-300 $7.4~\mu{\rm m}$ pitch $480\times640$ pixel Sony ICX424AL CCD camera giving a pixel scale at the focal plane of 0\farcs522. Optically, this is essentially identical to that employed at the Canada-France Hawaii Telescope (CFHT): a pupil mask of three 90 mm apertures, two with prisms separated by 170 mm from a third open aperture, all of which are equidistant from the optical axis. A diagram of the aperture geometry and resultant image spots is provided in Figure~\ref{figure_d14_geometry}. The USB-based Moravian camera can take 1 ms exposures approaching a 5 Hz duty cycle controlled with the manufacturer-supplied software running on a PC, although a separate software written in the C language was used to increase framerate. Some digital corruption of images was also associated with long USB extender communication cables, which restricted the position of the instrument on the roof.

The second DIMM, introduced in 2012, employs a Moravian Instruments G1-2000 camera essentially identical to the first, but with a $4.4~\mu{\rm m}$ pixel $1024\times1024$ Sony ICX274AK detector.  Otherwise, it is the same as to the Cerro Tololo Inter-American Observatory (CTIO) DIMM: two apertures of 80 mm, 174 mm apart. This DIMM uses a 25 cm (10 inch) Meade LX200 ACF 2500 mm focal-length Ritchey-Chr\'etien telescope feeding a combined MASS/DIMM unit surplus from the Thirty Meter Telescope (TMT) site-testing campaign - the same unit deployed on Mauna Kea. Precise optical alignment with MASS was obtained in the lab using a laser diode, following the procedures outlined in \cite{Kornilov2004}, yielding a MASS magnification factor of 14.52. With the Moravian camera used here, the DIMM provides a pixel scale of 0\farcs443.  This ``10-inch MASS/DIMM" (abbreviated MASS/D10) is operated with a Stealth Linux-based PC contained in an insulated box on the roof, with communication via ethernet down to a separate control PC in the warmroom.

Instruments were tested at -35C in an environmental chamber prior to deployment. It was found that self-heating by electronics was sufficient to keep the D14 camera and the MASS/D10 unit operational under all but the coldest conditions experienced. Cooling fans were de-activated and all possible locations where ice crystals (``diamond dust") could ingress were sealed.  Nevertheless, at times frost buildup did occur, and a webcam in a weather-proof enclosure aided in monitoring those situations. There was also some variation in focus of the Meade OTAs with temperature, requiring occasional manual adjustment, perhaps once or twice per week.

\subsection{Polaris as Target}

All observations used Polaris as the target. This is bright enough ($V=2$ mag) for easy acquisition and obtaining good focus even under thin cloud; it is also a multiple system, with Polaris A having a fainter companion Polaris B ($V=8$ mag, $\rho_{\rm AB}=18\farcs2$) which can allow a natural check on instrument sensitivity that will be discussed later.  A useful metric for when observations could reliably commence were hourly visual estimates of cloud cover at the Eureka weather station. Conditions reported as ``Clear", ``Mainly Clear" or ``Ice Crystals" in Eureka usually indicated skies were sufficiently clear to warrant beginning observations at PEARL - $V$-band extinctions reliably less than 2 magnitudes, typically what are considered ``spectroscopic" or better conditions \citep{Steinbring2012a}. Instruments were aligned visually; when contrast against the sky was sufficient. The Moon is always low at these latitudes, and did not interfere with data taking. So this was effectively a restriction to starting at Sun elevations below $-8$\arcdeg, which is most (but not all) of the time in October through February at PEARL. Even when the Sun came above $-8$\arcdeg, datataking might not necessarily be interrupted, but if so Polaris would not be reacquired until the Sun set again.

\subsection{Placement and Operation of Telescopes}

\begin{figure}
\plotone{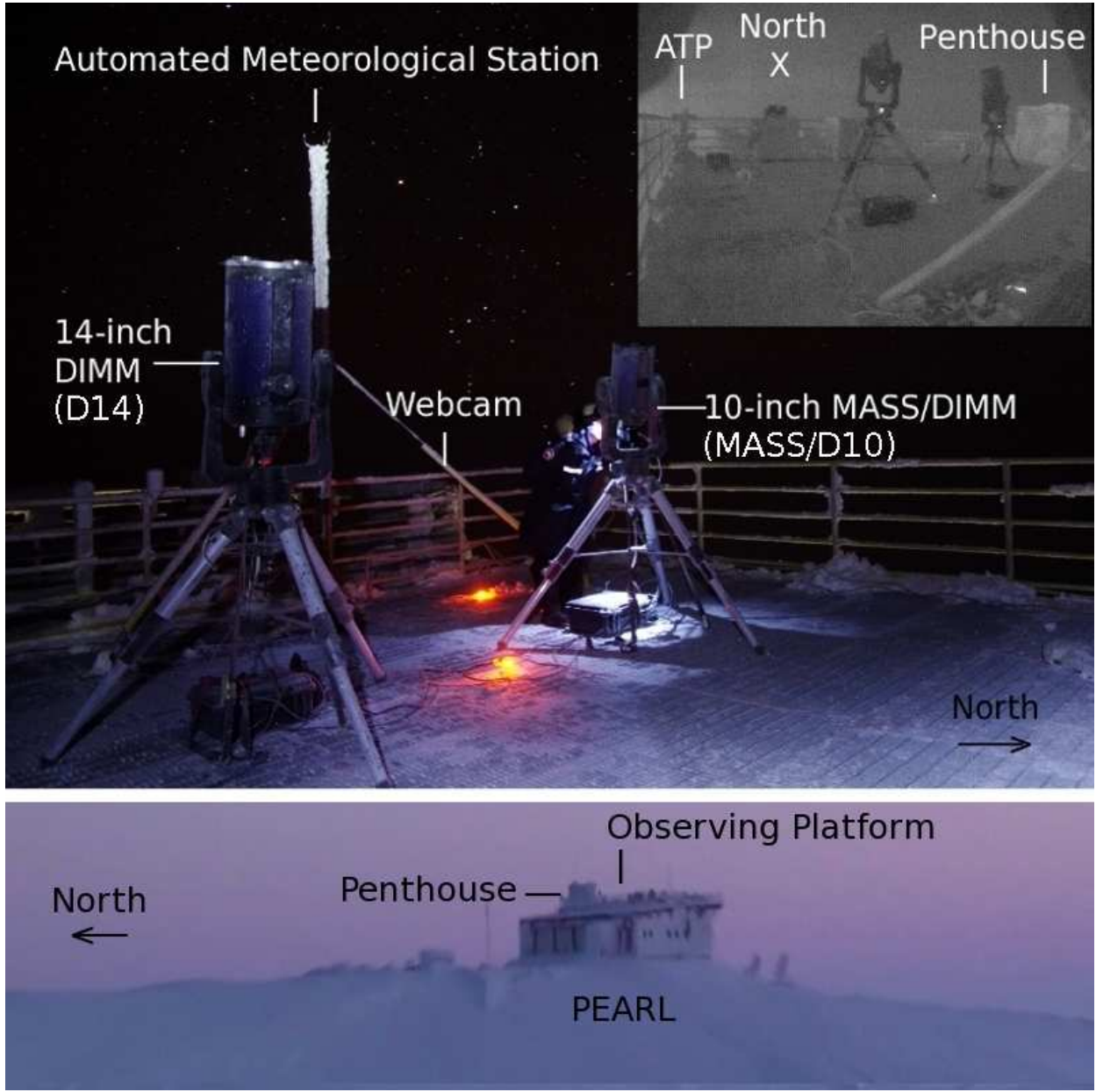}
\caption{D14 and MASS/D10 as deployed on the PEARL rooftop observing platform in February 2012. Inset is a view from the webcam, looking due north towards the location of the ATP. Below is PEARL seen from the northwest.}
\label{figure_telescopes_building}
\end{figure}

D14 and MASS/D10 were deployed on sturdy portable tripods, raising the entrance apertures approximately 2 m above the observing platform, roughly 8 m elevation from the surrounding terrain (see Figure~\ref{figure_telescopes_building}). As Polaris was used for all observations, the orientations of the telescopes were always the same to within 1\arcdeg~on the sky, and 10\arcdeg~from vertical. At first the stock Meade mounts were used, minimally modified: grease in the drives was replaced with some rated for operation at -40C, and spare drives were on hand. Although this was usually sufficient for remote operation from within the warmroom using a guider telescope, occasional loss of operability of the motor drives resulted in some times with low observing efficiency. As these were altitude-azimuth mounts, they also resulted in a consistent East-West alignment of the longest D14 sub-aperture baseline (and D10 baseline) on the sky. 

Telescopes were purposefully operated in the open, without enclosures. In 2011 February D14 was set up at the southwestern edge of the observing platform, to avoid being in the lee of the penthouse. In 2012 February D14 and MASS/D10 were set up side by side on the observing platform, towards its western edge, here still far away from the penthouse and air handling vents on the eastern side of the building, but closer to the northern edge of the platform. MASS/D10 was re-deployed in the same location here in October 2012, and then moved to the most northern edge of the platform for November 2012, near to the location where the ATP was deployed previously \citep{Hickson2010}. All of these positions are within 5 m of the mast of the autonomous meteorological station, aligned along a north-south line.

An issue was vibration due to flexing of the roof and from wind buffeting the telescopes. Some data are likely corrupted when observers were on the roof, with elongated images providing poor stellar centroids. These periods have not been deleted, but they are a small fraction, typically less than a few minutes per hour.  It was found that alignment of the telescopes on the target could not be obtained reliably when winds approached $8~{\rm m}~{\rm s}^{-1}$. The restriction was more serious for MASS/D10, as the MASS instrument field of view is smaller than for D14, and guiding could be lost more easily. Therefore, a uniform $8~{\rm m}~{\rm s}^{-1}$ shutdown wind speed was enforced, allowing both telescopes to be either taken down or covered before winds reached $10~{\rm m}~{\rm s}^{-1}$. This also avoided an unsafe condition for observers, as beyond that upper wind speed nobody is permitted to be alone on the PEARL roof.

\subsection{Image Quality, Signal-to-Noise Ratio, and Framerate}

Under perfectly clear (``photometric") skies and sharp focus, D14 and D10 data could be taken with 1 ms exposures, the shortest possible with the cameras.  This provided a stellar flux through a single D14 aperture, for example, of $F_{\rm D10}\approx8300~{\rm ADU}$, or a per-frame signal-to-noise ratio of $S/N={{G F_{\rm D10}}\over{R}}\approx400$, for detector gain $G=0.334~{\rm e-}~{\rm ADU}^{-1}$ and readout noise of $R=7~{\rm e}^{\rm -}$. To maintain this minimum $S/N$ under thin cloud, when $V$-band extinction went beyond 0.5 mag, a continuous sequence of 3 ms exposures was used instead. This allowed centroids under all conditions except thick cloud. For D14 in 2011 these 3 ms exposures were interleaved with 6 ms exposures, allowing monitoring of bias due to a non-zero exposure time. A difficulty with this procedure is that it trades off the number of data samples per minute - an equal division effectively cutting those in half - with the accordant cost to statistical sampling. Longer averages are undesirable as they could hinder correlation with meteorological data, which are averaged on 60 second intervals, as is MASS. It was found that a consistent 5 Hz framerate (300 samples per minute) for both D14 and D10 could be achieved within the limitations of communications bandwidth and computer speed if only subsections of the detectors containing just the star images ($200\times200$ pixels) were read out and stored on the harddisks.

Provided skies had only thin cloud, typical fluxes in the outermost MASS aperture (`D') were reliably over $F_{\rm D}=200$ counts. About 22\% of the time it fell below $F_{\rm D}=100$ counts, under the $S/N$ limit recommended in \cite{Kornilov2007}, and observations were excluded. Some of that time may also be affected by frost or ice-crystal buildup on the entrance aperture. But when icing occured, it appeared from inspection to be uniform over the entrance pupil, and the relative difference in flux between aperture D and B, $(F_{\rm D}-F_{\rm B})/F_{\rm D}$, remained stable, not varying by more than 5\%. None of the observations suffer from $S/N$ poorer than 100, i.e. errors in flux measurement in aperture D of $\delta F_{\rm D}<0.01$. Also, under the conditions when MASS operated, with sun elevations below $-8$\arcdeg, the relative background of sky $B_{\rm D}$ versus aperture D flux was always less than 2\%. Kornilov et al. recommended $B_{\rm D}/F_{\rm D}<0.03$.

Image quality was monitored during observations. For D14, the PSF of the stellar images through the prisms were slightly elongated (along each baseline towards the open aperture) with an ellipticity close to $0.1$. Shown in Figure~\ref{figure_d14_geometry} is a single $200\times200$ pixel frame taken with the instrument precisely focussed during some of the better seeing recorded.  A quadratic stretch has been applied to the image, with star images truncated at half-peak flux. The open aperture was helpful in guaging the delivered optical quality of the system, because broadening of this near-Gaussian PSF beyond its nominal diffraction limit of ${\rm FHWM}_\circ=1.028{\lambda\over{d}}$, or $1\farcs18$ where wavelength $\lambda = 0.5~\mu{\rm m}$ and aperture diameter $d=90~{\rm mm}$, was attributable primarily to defocus. As total flux must be conserved, Strehl ratio $S$ (image peak relative to perfect diffraction) of the stellar image then degrades approximately as $({\rm FWHM}_\circ/{\rm FWHM})^2$. Those with $S<0.6$, when the elongated PSFs may be poorer $S=0.5$, were rejected, which eliminated only data taken in October 2012. To minimize the effect of instrument aberration in other data, just the two displacements (in projected longitudinal and transverse components) between a prism and open aperture were used in later analysis.

\begin{figure}
\plotone{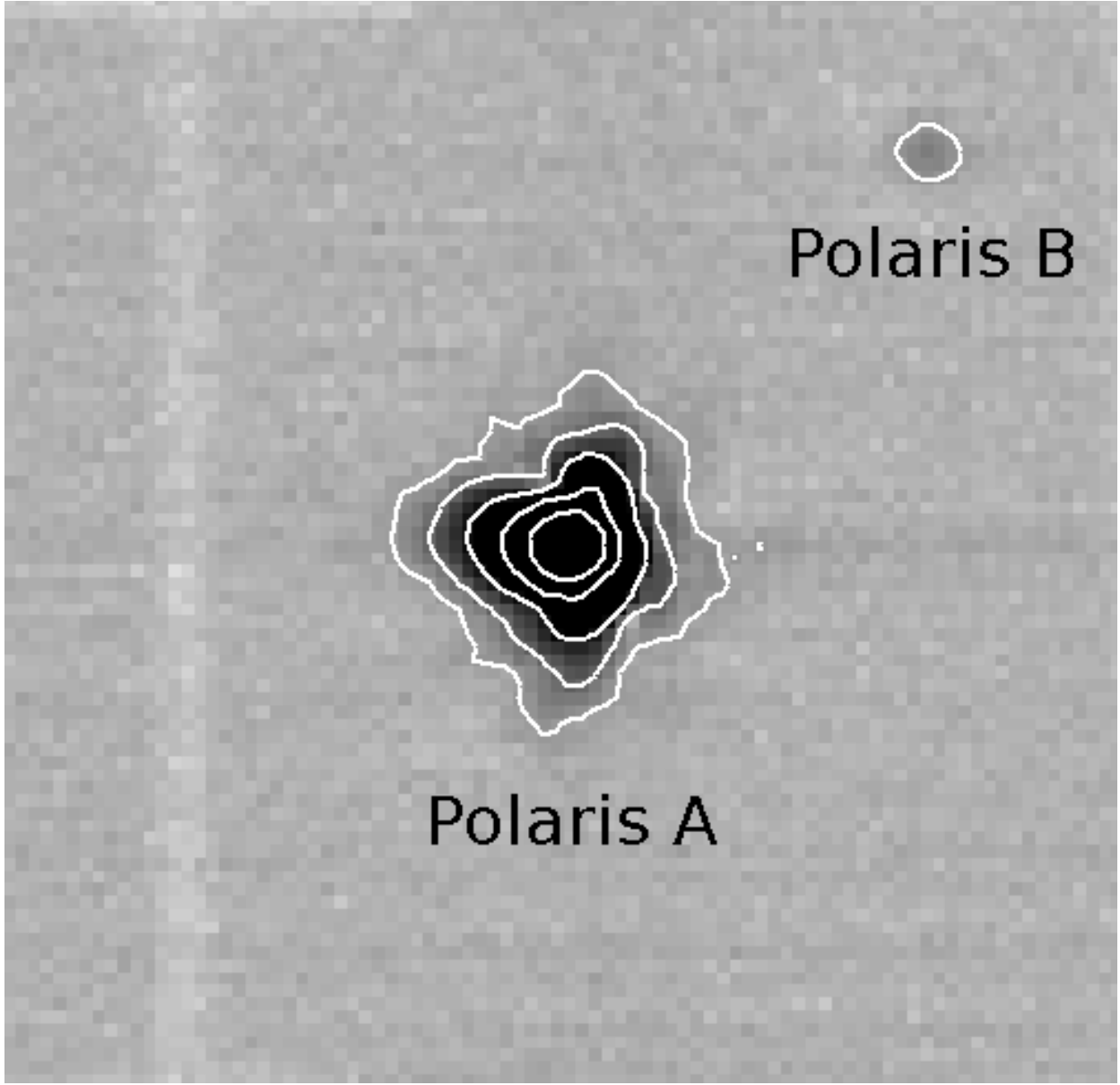}
\caption{Images of Polaris A and B obtained from one aperture of D10 by combining $2\times10^4$ consecutive frames - a total integration of 60 s. The D10 diffraction limit at 0.5 $\mu{\rm m}$ has ${\rm FWHM}=3.00$ pixels; image quality is good, with the first diffraction ring visible for Polaris A in this logarithmic stretch: contours overplotted at 0.2, 0.05, 0.01, $2\times10^{-3}$, and $5\times10^{-4}$ peak flux. Polaris B appears at a separation of $\rho_{\rm AB}=18\farcs1\pm0\farcs2$; the subframe is 100 pixels on a side.}
\label{figure_polaris}
\end{figure}

A further check on instrument performance came from stacking images to verify the faint companion Polaris B was visible. \footnote{Polaris A is also binary, although Ab has $V=9$ and $\rho_{\rm AaAb}=0\farcs18$ in {\it Hubble Space Telescope} imaging \citep{Evans2008}; it is undetectable in these data.} This was particularily effective when D10 operated on the equatorial mount, although digitally counter-rotating frames also worked. Figure~\ref{figure_polaris} shows the sum of $2\times10^4$ co-registered $100\times100$ pixel subframes from one sub-aperture of D10, all in consecutive sequence from 17 October 2012.  Other observational periods are similar. (Finding DIMM sub-pixel image centroids will be discussed in Section~\ref{observations}.) This represents $60~{\rm s}$ of integration obtained over the course of 67 minutes of continuous operation. Polaris B is found to be $780\times$ fainter than Polaris A, with a PSF FWHM differing by less than 1\%.  This FWHM of 1\farcs57 is relative to a nominal diffraction limit of ${\rm FHWM}_\circ = 1\farcs33$, which implies $S\approx0.72$ without strong field-dependent aberrations. All D10 data maintained $S>0.6$, as recommended by \cite{Kornilov2007} to avoid high-altitude turbulence bias during MASS/DIMM operation.

\subsection{Sky Clarity, Wind, and Sampling Efficiency}

With the instruments well controlled, the dominant restrictions to observing efficiency remain the combination of sky clarity, sky brightness and wind speed. It will be shown later that seeing and ground winds are correlated in these data, with best results tending to occur when winds were light. Previous studies have already shown these to presage clearer skies at PEARL. Observing efficiency was sufficiently good, however, that this should not constitute a strong bias. For example, in February 2012 there were 444 hours when observers were at PEARL.  Of these ``observable hours" 290 hr were sufficiently dark to allow alignment on Polaris, and of those $(290~{\rm hr}-178~{\rm hr})=112~{\rm hr}$ were cloudy, according to visual inspection from Eureka. This is a clear-sky fraction of $178~{\rm hr}/290~{\rm hr}=61$\% during which 153 hr was not windier than $8~{\rm m}~{\rm s}^{-1}$.  During this time 115 hr of datataking took place with D10, for an observing efficiency of $115~{\rm hr}/153~{\rm hr}=75$\%.  Efficiencies are roughly half that for D14, still sufficient for a useful sample.

\begin{figure}
\plotone{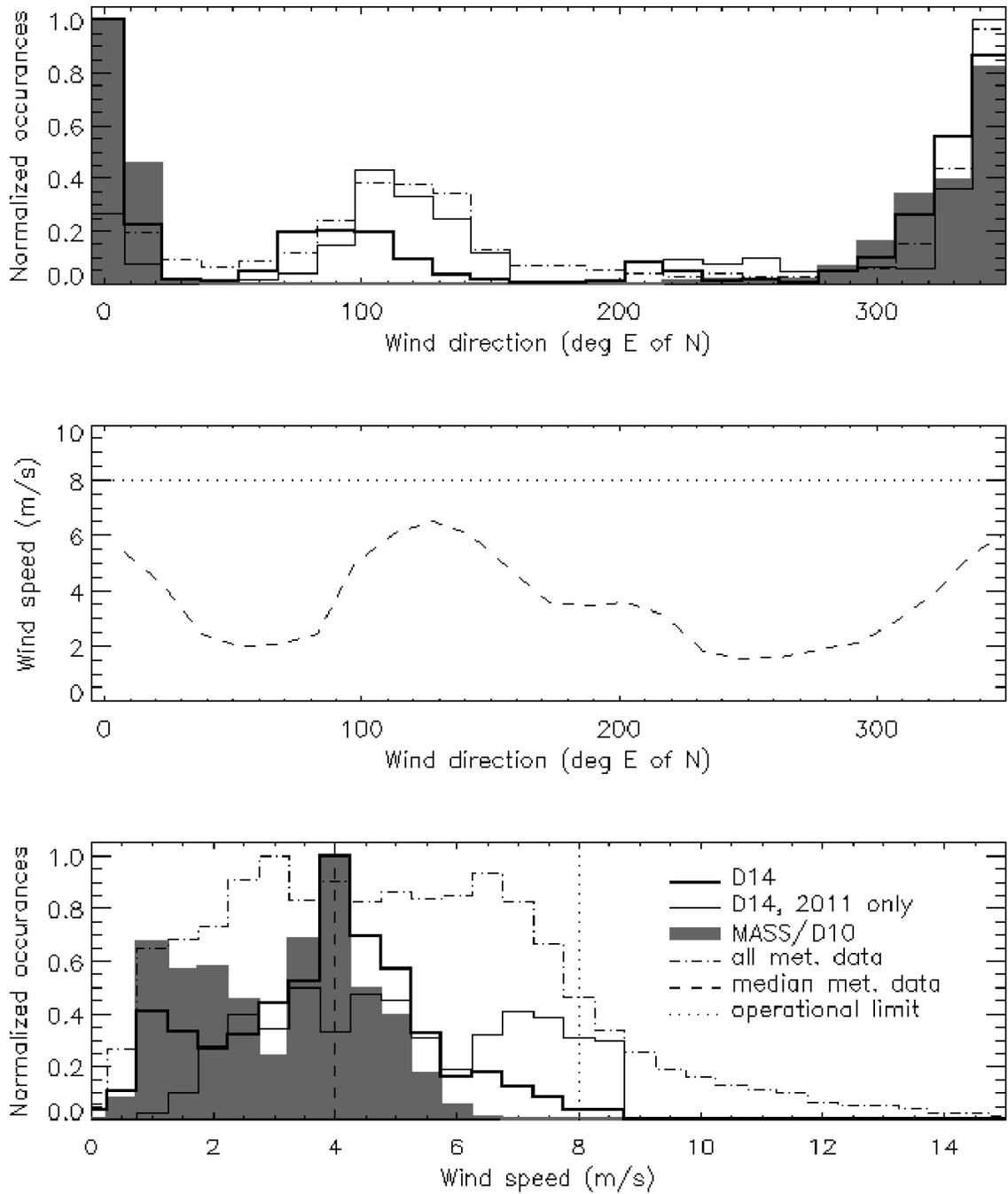}
\caption{Distributions of wind direction at PEARL (top); only times when D14 data were taken (thick black), those during 2011 (thin black) and MASS/D10 (grey shading). The median wind speed as a function of wind direction is shown (middle), as well as a histogram as a function of wind speed (bottom). MASS/D10 data tend to sample better conditions, with less windy weather than D14 and not in the lee of the PEARL penthouse.}
\label{figure_windspeed_distributions}
\end{figure}

Figure~\ref{figure_windspeed_distributions} shows the distribution of wind direction and speed for all D14 and MASS/D10 observations where PEARL meteorological measurements are available. Good weather at PEARL occurs typically when winds are from the north or northwest, with cloudy weather often approaching from the south. Not surprisingly, wind directions to the north are better sampled. Note that cloudy conditions in Eureka, associated with storms, are known to be correlated with strong winds \citep{Steinbring2012a}. In effect, when visual inspection in Eureka reported skies ``Mostly Cloudy", ``Cloudy" or ``Indeterminant" (usually snow) it is unlikely that data could be taken at PEARL.

A strong peak in the sampling distribution at $4~{\rm m}~{\rm s}^{-1}$, also the median wind speed for PEARL, may be attributed to observers beginning observations whenever winds dropped below this speed. The result is a tendency towards observations very near this speed, as these could cease if wind speeds were to rise quickly afterwords, or perhaps stop for technical faults unrelated to weather. 

As mentioned previously, some loss in observing efficiency was due to frost buildup obscuring the telescope apertures. When this happened, the D14 aperture mask windows were wiped with a lens-cleaning tissue and methanol. This was less successful for D10 as the entirety of its entrance aperture corrector plate must be clear to allow datataking with MASS. Either frost, or a combination of that with cloud, sometimes stopped observations. This is consistent with the distribution in Figure~\ref{figure_windspeed_distributions}, the tendency was that D14 was somewhat more likely to be operating during windier, cloudier weather than D10. Some improvement came in 2012 with a better method to remove frost from the MASS/D10 window using occasional use of a blower.

The lack of observations in strong winds, particularly approaching $8~{\rm m}~{\rm s}^{-1}$ is also evident in Figure~\ref{figure_windspeed_distributions}. Some general improvement in observing efficiency for MASS/D10 came in October 2012 when a more robust Astro-Physics GTO3600 German-equatorial mount was employed. This also relaxed the restriction to commencing observations when winds were light.

\section{Data and Reductions}\label{observations}

Observations with either D14 or MASS/D10, or both, were obtained whenever it was possible to track Polaris: starting when Sun elevations were below $-8$\arcdeg~with sustained wind speeds at or below $4~{\rm m}~{\rm s}^{-1}$ at PEARL, and continuing while skies were sufficiently clear and winds remained under $8~{\rm m}~{\rm s}^{-1}$. A journal of observations appears in Table~\ref{journal_of_observations}.

\subsection{D14 and D10}

\begin{figure}
\plotone{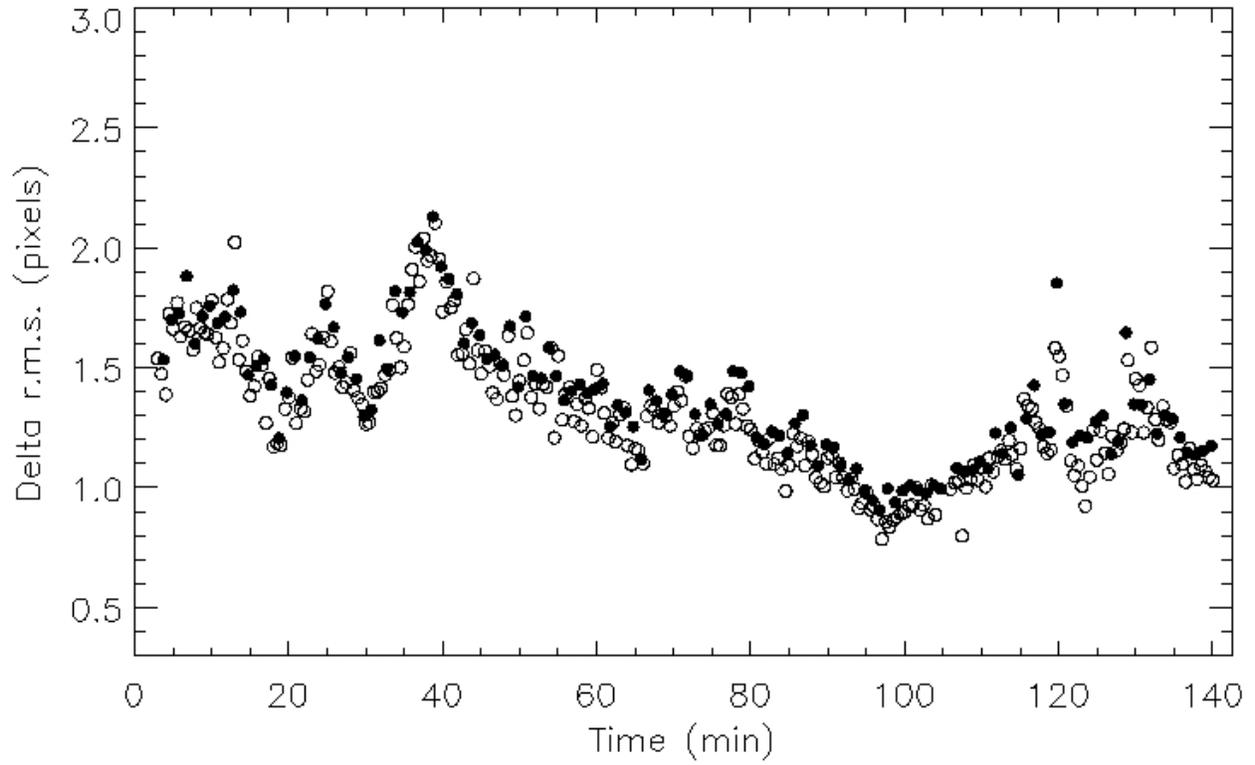}
\caption{Comparison of D14 centroiding for two different techniques; averages of r.m.s. displacements during sampling periods using DAOFIND (60 s, filled circles) and SExtractor (30 s, open circles).}
\label{figure_centroids_zoomed_in}
\end{figure}

DIMM reductions followed closely those described in \cite{Tokovinin2002} using the ``Z-tilt" conversion coefficients advocated in their equation 8. Row and column displacements of each pair of star images on the detector were obtained using two techniques for D14: first based on the SExtractor software, and a second using an IDL implementation of DAOFIND in IRAF. For both, a synthetic circular aperture close to the optimally small value of radius $r=1.22{\lambda\over{d}}$ was used: 3 pixels for D14 and 4 pixels for D10. Only the SExtractor method was used for D14 data in 2011. These yielded similar results, although the latter was somewhat more sensitive to image quality, tending to reject frames with broader PSFs, and is likely to be more conservative, i.e., predicting worse seeing (see Figure~\ref{figure_centroids_zoomed_in}). This was the preferred method, and adopted for D10. Each method provided four independent displacements for the D14 (just two for D10) and the variances of which were then averaged over intervals of either 30 s (D14) or 60 s (D10).

Standard calibrations were applied to report final seeing estimates: a plate-scale conversion obtained from observing a bright binary star of known separation, debias of extraneous variance due to detector noise, and correction to zero exposure time. Detector noise was subtracted following the method of equation 9 in \cite{Tokovinin2002}. At the minimum $S/N=400$ and using an $r=4$ pixel synthetic aperture this amounts to 0\farcs05 (r.m.s.) for D10, and with D14 at $r=3$ it is 0\farcs02. Note that these are maxima, with noise decreasing for better $S/N$ data. Two different schemes were followed to scale to zero exposure time. In February 2011, for D14 data, 3 ms exposures were interleaved with 6 ms exposures, allowing the double-exposure reduction discussed in \cite{Tokovinin2002}. This corrective factor (always in the sense of reporting worse seeing) depends on wind speed and the distribution of turbulent layers in the atmosphere, and an iterative procedure accounted for that variation - in effect, smoothing it over time. Although this correction was sometimes over 30\% during the poorest observed seeing (possibly influenced by extreme wind conditions), when sub-arcsecond it was usually under 10\%. In 2012 a uniform 3 ms exposure time was used for both D14 and D10 datataking, which was corrected by using available sequences of 1 ms exposures; an average 7\% increase in seeing was applied throughout. Even if this were an underestimate of the zero-exposure-time correction when seeing is poorer than 1\arcsec, it should not affect the median significantly. A geometrical correction for airmass was also applied, although it is small for Polaris at this latitude ($\sec z<1.02$).

\subsection{Combination with MASS}

MASS data were taken whenever D10 operated, and usually preceded it because the alignment of both instruments was obtained through the MASS viewfinder. A 60 second integration time was employed, producing a turbulence profile corrected to zenith using the reduction software Turbina \citep{Kornilov2003} Version 2.047, including correction for strong-scintillation ``overshoot" discussed in \cite{Kornilov2007}. The formal errors in model fits for MASS profiles were all $\chi^2<100$, as recommended for data quality in Kornilov et al. Estimates of the coherence time and isoplanatic angle are also reported, which will be the subject of another paper, only seeing is discussed here. Turbina integrates the $C_n^2$ profile in 6 layers (500 m, 1 km, 2 km, 4 km, 8 km, and 16 km and up) above ground level, providing an estimate of the FA seeing - which for these data has a median of 0\farcs50, and a mode of 0\farcs23.

\begin{figure}
\plotonehalf{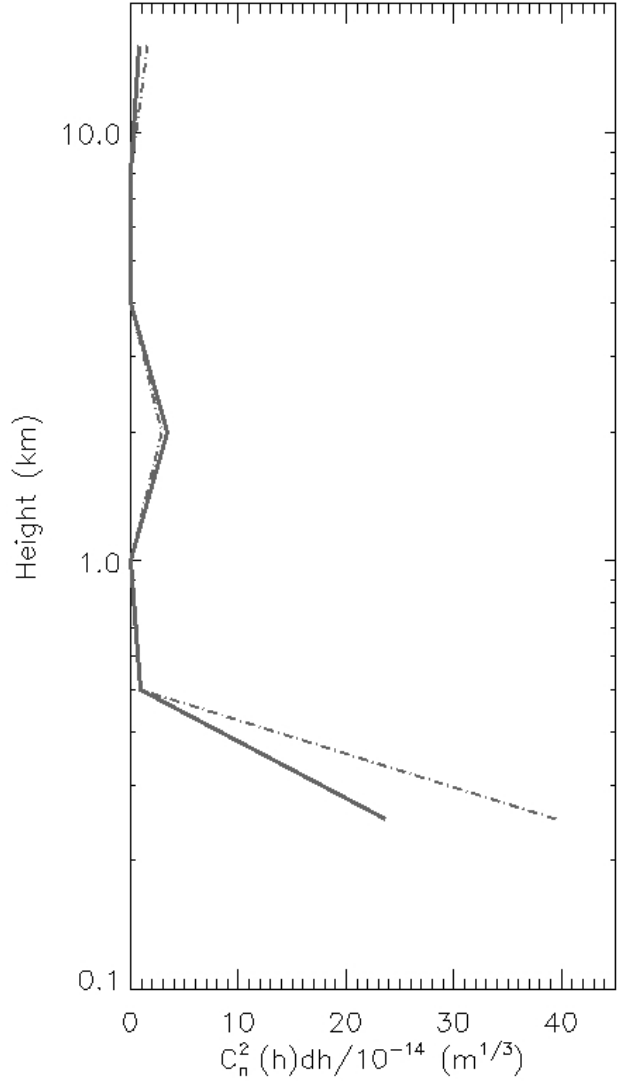}
\caption{Median profiles of $C_n^2$ from MASS: six layers from 500 m to 16 km, combined with a ground-layer component computed from coincident D10 seeing; thick grey curve is for times when skies were ``Mainly Clear" or better, and dot-dashed curve indicates all data. Height is given relative to PEARL, which is at an elevation of 610 m above sea level.}
\label{figure_results_profiles}
\end{figure}

The MASS instrument is not sensitive to turbulence below 500 m, but this can be estimated from D10, which is sensitive to the integrated turbulence along the same optical path. The component in the ground layer (GL) can then be estimated via \cite{Skidmore2009}:
$$\epsilon_{\rm GL} = {{(\epsilon_{\rm DIMM} - \epsilon_{\rm MASS})}\over{|{\epsilon_{\rm DIMM} - \epsilon_{\rm MASS}|}}}\times{|{\epsilon_{\rm DIMM}^{5/3} - \epsilon_{\rm MASS}^{5/3}|}^{3/5}}. \eqno(1)$$
A plot of the resulting 7-layer profile is shown in Figure~\ref{figure_results_profiles}. Only D10 data which were concurrent with MASS were used to calculate a ground-layer component. This has a median of 0\farcs61, that is, an integrated seeing of $(0\farcs61^{5/3} + 0\farcs50^{5/3})^{3/5} = 0\farcs84$. Restricting to times when skies were reported ``Mainly Clear" or better reduces the median of $\epsilon_{\rm GL}$ to 0\farcs44. The effect is much less for the FA seeing, only 20\% difference in $C_n^2(h)$ (primarily at $h=2~{\rm km}$, see Figure~\ref{figure_results_profiles}), which suggests that uncertainty in the integrated seeing is dominated by variation in the ground layer. 

\begin{figure}
\plotonenarrow{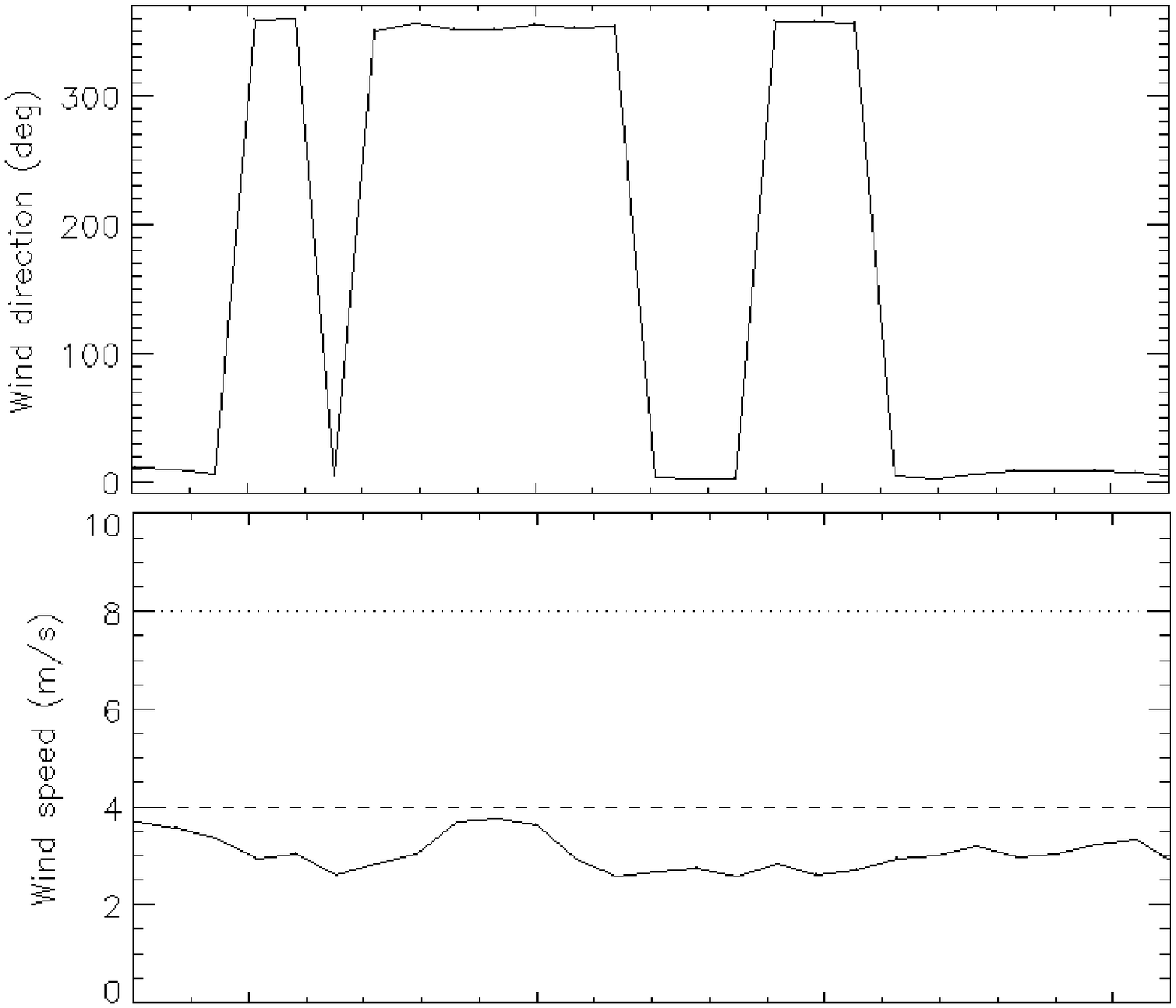}\\
\plotonenarrow{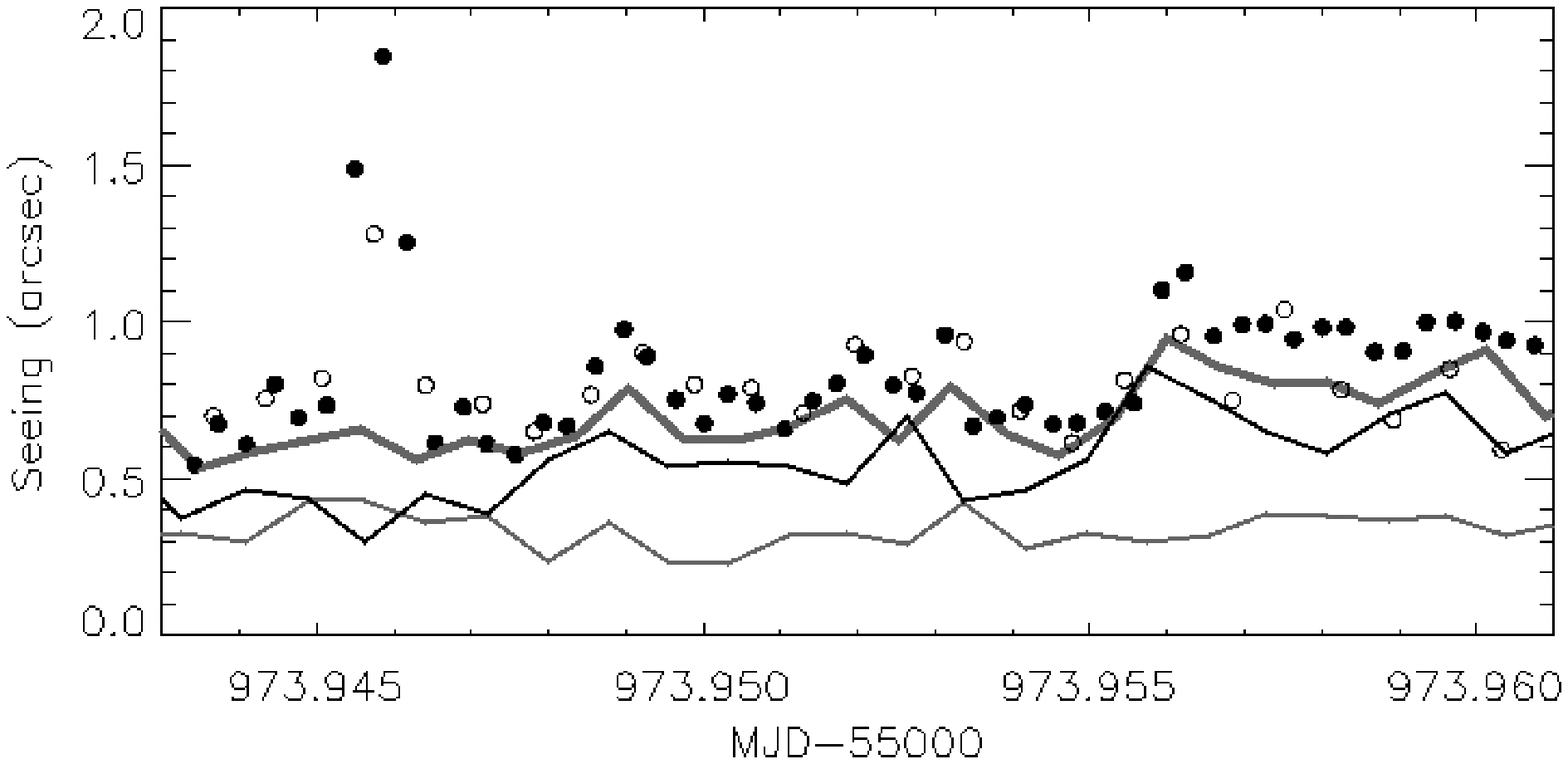}
\caption{A sample period when all instruments were in operation: PEARL wind direction and speed, D14 seeing using DAOFIND (filled circles) and SExtractor (open circles); D10 (thick grey lines), MASS (thin grey), and GL (thin black). In general there is good agreement, although the MASS/D10 does not seem to detect the ``seeing bubble" near ${\rm MJD}-55000=973.946$. During that brief ($\sim 2~{\rm min}$) episode the wind was near $3~{\rm m}~{\rm s}^{-1}$, between north and north-northwest.}
\label{figure_results_overlap}
\end{figure}

As an example, a sample of one short period which included observations with all instruments is shown in Figure~\ref{figure_results_overlap}. Here, D14 data are shown sub-sampled to 30 second intervals, and the alternate reduction method for D14 (using SExtractor) is overplotted as open circles. The D10 data are shown connected with a thick grey line; MASS data connected with a thin grey line.  Although not identical in detail, there are general similarities between D14, D10 and MASS data. This is not a surprise for the latter two, as they were implemented on the same telescope.  Differences in the timing of samples may account for some discrepancy between the D14 and D10, perhaps even the brief ``seeing bubble" which occured at ${\rm MJD}-55000=973.946$. That appears to have been a local phenomenon, near the ground, as it was not detected by MASS.

All seeing data are plotted in Figure~\ref{figure_weather_and_seeing_results}. Also shown here are Sun elevation and meteorological data for the first two observing runs (PEARL meteorological data records are not available for 2012). This helps illustrate the observing restrictions mentioned above, with periods of cloudy skies and high winds not being sampled. 

Some of the best seeing was recorded in November 2012. This was when D10 was situated at the most northern edge of the observing platform, and possibly least affected by the building itself.  During that time the median D10 seeing was 0\farcs64, although over the previous two deployments it had a median of 0\farcs86. It is perhaps a little surprising that some of the poorest seeing recorded occurred during periods of clear skies, e.g. near ${\rm MJD}-56000=219$. Note that instances of strong thermal inversion can be windy at PEARL, for example, the period near ${\rm MJD}-55000=974$ discussed above. Possible bias related to local wind direction and speed will be discussed in more detail in the next section.

\section{Analysis and Discussion}\label{analysis}

Although only limited MASS data have been collected so far - about 118 hours - these do consistently show weak FA turbulence, with a dominant layer near 2 km. As local and boundary-layer effects may play a more important role in the integrated seeing at PEARL, further analysis was carried out with regard to seeing distributions with ground wind speed and direction. From Figure~\ref{figure_windspeed_distributions} it can be seen that samples with the D14 are representative of the overall distribution, and MASS/D10 data (although there are significantly more of these than with D14) tend to better sample winds near the median wind speed or less.

Figure~\ref{figure_seeing_correlations} shows the correlation of seeing with ground wind direction and speed at PEARL for all DIMM and MASS data where simultaneous PEARL meteorological data are available. Medians of D14 data in 15\arcdeg~and $0.5~{\rm m}~{\rm s}^{-1}$ bins are overplotted, with error bars indicating the formal 1-$\sigma$ deviations in each bin; open circles are for 2011 D14 data only. A thick grey curve indicates medians of D10 data. Note that in the bottom plot of wind speed, the medians are restricted to wind directions greater than 310\arcdeg~and less than 40\arcdeg~east of north, to match conditions when D10 data were taken.

Note that most samples were collected when winds were from the north, as expected, and poorer seeing tends to be associated with southerly winds - the prevailing direction during stormy winter weather at Eureka. The penthouse was always to the northeast of D14, but closer to east for MASS/D10.  

Despite these differences in deployment there is agreement between 2011 and 2012 D14 data when winds were from the north (and with D10 when winds are greater than $4~{\rm m}~{\rm s}^{-1}$). In the bottom plot there is no bin where these medians are discrepant by greater than the formal statistical scatter within a bin, and are typically within 10\%. This is reassuring, as the reduction methods differ: the 2011 D14 data reduction used SExtractor centroiding and interleaved exposures to determine the correction to zero exposure time, and the 2012 D14 data used DAOFIND and a fixed corrective factor. One would expect that failing to fully account for a finite exposure time would result in an optimistic estimate of seeing, and there is no evidence of that here.

There is evidently a trend towards poorer DIMM seeing with increasing wind speed, at least beyond $4~{\rm m}~{\rm s}^{-1}$.  Seeing in lighter winds than this are also worse, and more pronouncedly for D14. A thin black line indicates the medians for D14 after applying a further restriction that samples be taken under visually ``Mainly Clear" or better conditions. This would be a better match to times when data were obtained with the D10, which could not tolerate cloud as easily, and may in part explain the difference in results between the two. Best median seeing was recorded with D10 at wind speeds closer to $3~{\rm m}~{\rm s}^{-1}$, although it too indicates poor seeing when winds were calm.

It may be possible to further characterize the effects of ground winds on the seeing at PEARL. Following the discussion in \cite {Salmon2009} concerning dome seeing at CFHT, one might expect this to scale as $v^{6/5}$, simply an additive variance assuming a Kolmogorov spectrum of turbulence in a dominant layer (presumably near the ground). A curve which assumes $$\epsilon_v\propto v^{6/5} \eqno(2) $$ has been fit by a least-squares method for winds in excess of $4~{\rm m}~{\rm s}^{-1}$, and is overplotted in Figure~\ref{figure_seeing_correlations}. This gives a remarkably good fit above the median wind speed up to the limit of operations, with an intercept of $0.30\pm0.04$ arcsec. 

That this curve does not fit the data below the median ground wind speed could be explained by one or a combination of two factors: poorer FA seeing or local effects including the influence of the building itself. It may seem counterintuitive for low ground wind speeds to be correlated with poorer FA seeing, but this is possibly related to the thermal inversion, as it may also be weaker then. At the same time, with low ground wind speed, DIMM seeing measurements could be spoiled by a heat plume rising from the building (or even be related to the instrument itself, as D14 and D10 always pointed near zenith). The assumption of fully involved Kolmogorov turbulence may break down, with weak flow producing large eddies over the roof. However, there are not yet sufficient data here to differentiate between the two scenarios - poorer FA seeing or local seeing - as it would require sampling many occasions of complete calm with a strong inversion.

Histograms of all the data are shown in Figure~\ref{figure_seeing_distributions_model}. The results from ATP measurements - not simultaneous in time, but from a similar location - are also shown.  This curve is based on combining the distribution of ground-layer measurements with a uniform FA seeing of 0\farcs30.  Note how near the D10 cumulative distribution approaches this curve. Limiting the D14 dataset to times when winds were above the median wind speed also yields a result closer to the D10 result: $0\farcs72\pm0\farcs04$, with a 5\% uncertainty assumed from the fitting error using equation 2. A thin line indicates the distribution of D14 data, excluding those taken with winds below $4~{\rm m}~{\rm s}^{-1}$, which helps illustrate the influence of poor seeing when winds are calm.  

It is also interesting to consider what seeing distribution the simple model would predict based on the known wind speed distribution at PEARL, including when DIMM data were not collected.  This is shown as a dot-dashed curve obtained by convolving the distribution of wind speeds for all ``observable" times (during seeing observations or not) with the fit of equation 1 in Figure~\ref{figure_seeing_correlations}. This predicts the same modal seeing as measured with D10.  Also note how well this curve fits the best seeing of the D10 cumulative-distribution curve, yielding identical 20\%-ile seeing of $0\farcs53\pm0\farcs03$, also similar to the median seeing estimated for the ATP at 7 m elevation \citep{Hickson2013}. Part of this difference is due to the fact that the model incorporates only a fixed value for FA seeing, not a distribution. It is also possible that the ATP, having sensors somewhat higher than the aperture of either D14 or D10, is less corrupted by turbulence associated with the building itself.

\section{Summary and Conclusions}\label{summary}

Seeing measurements at PEARL obtained with DIMM and MASS have been presented, the first on a High Arctic coastal mountain site. Two instruments were deployed: a stand-alone DIMM and on a 14-inch telescope (D14) and one combining both DIMM and MASS on a 10-inch telescope (MASS/D10). Free-atmospheric seeing measured with MASS has a median of 0\farcs50 and mode of 0\farcs23, with a dominant layer at 2 km above the site which varies less than the ground layer. This is in reasonable agreement with previous estimates of the boundary layer seeing at the same site obtained with the ATP. Although not sufficient for robust statistics at the site, these new data are representative of typical weather during winter darkness, and do point to the occurrence of excellent conditions, even from the roof of the building. Seeing is best when winds are northerly, and not calm. D14 gives a median seeing better than 0\farcs85 under clear skies, which differs from the results for D10. The building itself is probably at least partly to blame for this discrepancy. A simple model of seeing as a function of wind speed was developed which helps clarify this. The model allows better agreement between the two, which can be explained naturally by their different sampling of wind conditions at PEARL. The median seeing at PEARL with D10 is $0\farcs72\pm0\farcs04$, with a 20\%-ile of $0\farcs53\pm0\farcs03$.  

To our knowledge, no other astronomical observatory site in Canada has demonstrated consistent sub-arcsecond seeing. And remarkably, these first results at PEARL are comparable to some of the best sites worldwide: median nighttime DIMM seeing of 0\farcs79 and MASS seeing of 0\farcs50 at CTIO \citep{Els2009a} and medians of 0\farcs75 (DIMM) and 0\farcs33 (MASS) at Mauna Kea 13N \citep{Skidmore2009}.

More data are needed to fully characterize the statistics of seeing at the PEARL site during polar darkness. Current plans are to automate MASS/D10 and obtain data over at least one complete winter season. This will provide a robust estimate of FA seeing by sampling a full range of thermal inversion strengths. Ground wind conditions will also be more representative, as commencement of each operational period can cease to be dependent on wind speed, and robotic operation will eliminate the requirement of sky-brightness and clarity conditions suitable for visual alignment by ``eye." To ensure consistent application of a bias correction due to finite exposure time it would be best to implement interleaved DIMM exposures of 3 ms and 6 ms. The instrument should be situated at the most northern edge of the PEARL observing platform, although this could still be insufficient to remove doubt concerning corruption of seeing by the building itself. Operation from an isolated tower of 6-m height or greater, away from the building would be necessary, and that is being developed.

\acknowledgements

The MASS instrument was loaned from CFHT, and we thank in particular observatory staff Derrick Salmon and Marc Baril for their assistance. Ivan Wevers and John Pazder helped prepare and test the instruments in the NRC-Herzberg cold chamber.  We are grateful for the wonderful support of the NRC-Herzberg and UofT machine-shop staff.  We thank Jerome Maire of the Dunlap Institute for his assistance during the October/November 2012 campaign. It is also our pleasure to acknowledge the Canadian Network for the Detection of Climate Change, particularly Pierre Fogal, James Drummond and technician Alexei Khmel for support with operation of the instruments, communications, and for providing the PEARL meteorological data.  ES, MM-B and WN are grateful to Environment Canada and the staff of the Eureka weather station for their hospitality during our ``observing runs."  We thank Andrei Tokovinin for many helpful comments on the original manuscript. This research was funded through NRC Canada and grants from the Natural Sciences and Engineering Research Council of Canada.

\newpage

\begin{deluxetable}{lccc}
\tablecaption{Journal of Observations\tablenotemark{a}\label{journal_of_observations}}
\tablewidth{0pt}
\tabletypesize{\small}
\tablehead{\colhead{} &\colhead{14~inch} &\multicolumn{2}{c}{10~inch}\\
\cline{3-4}
\colhead{Dates} &\colhead{DIMM (D14)} &\colhead{MASS} &\colhead{DIMM (D10)}}
\startdata
2011, 8 - 18 Feb     &~400    &\nodata &\nodata\\
2012, 10 - 22 Feb    &1439    &~735    &~914\\
2012, 12 - 22 Oct    &\nodata &3318    &3096\\
2012, 25 Oct - 7 Nov &\nodata &3059    &3772\\
\cline{1-4}
Total                &1839    &7112    &7782
\enddata
\tablenotetext{a}{In samples of 60 s duration.}
\end{deluxetable}

\newpage

\newpage

\newpage

\newpage

\newpage

\newpage

\newpage

\newpage

\newpage

\begin{figure}
\plotone{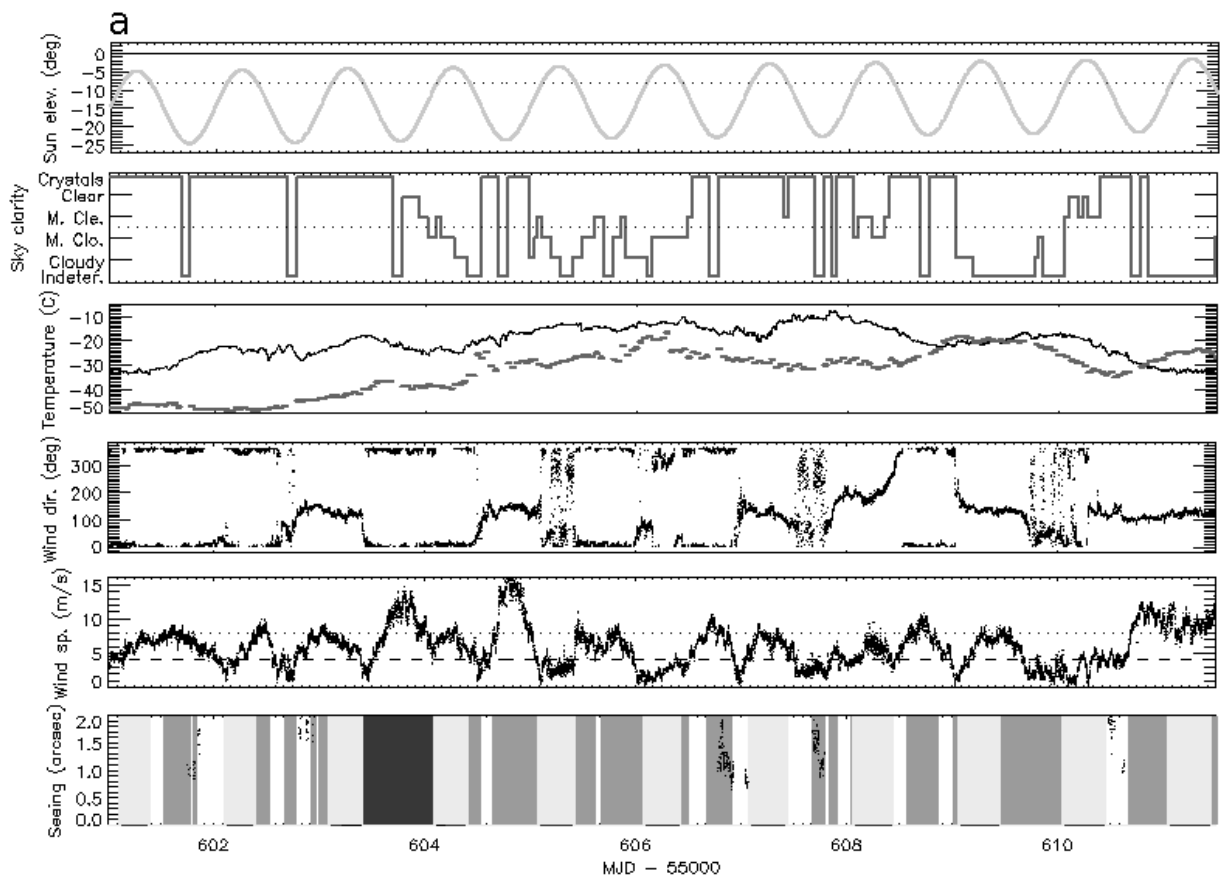}
\caption{Seeing from D14 (black points), D10 (grey points), and MASS (grey lines) for the four roughly one-week-long observing runs. For the first two runs, February 2011 (a) and February 2012 (b) the corresponding Sun elevation and meteorological data are plotted; visual sky clarity (at Eureka), temperature (grey: Eureka; black: PEARL), and wind direction and speed (at PEARL) with median wind speed indicated by a dashed line. Conditions worse than indicated by horizontal dotted lines precluded reliable measurements; these periods are ``greyed-out" in the seeing plots.  Note that no PEARL ground wind speeds are available for October/November 2012 (c and d). Light shading indicates bright time. No observers were present during ``blacked-out" periods.}
\label{figure_weather_and_seeing_results}
\end{figure}

\newpage

\setcounter{figure}{8}

\begin{figure}
\plotone{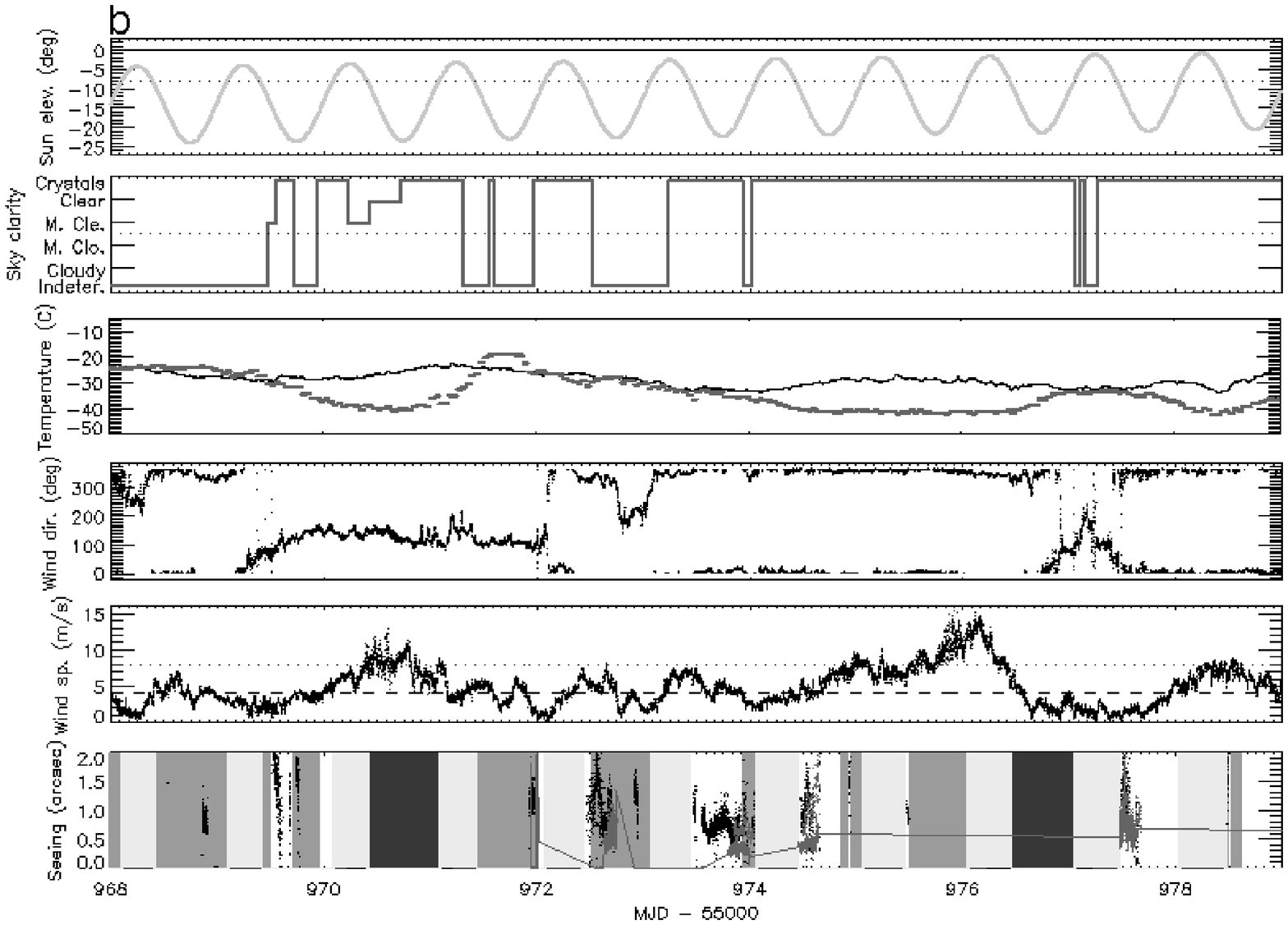}
\caption[]{Continued.}
\end{figure}

\newpage

\setcounter{figure}{8}

\begin{figure}
\plotone{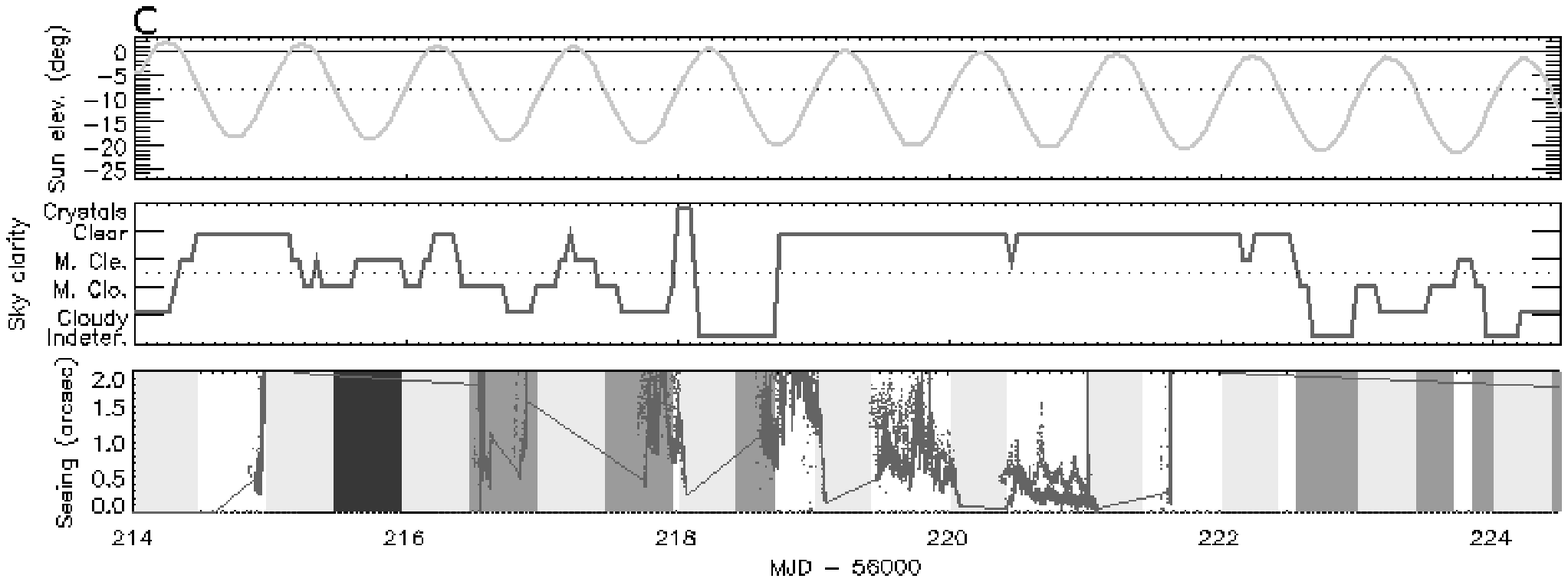}\\
\plotone{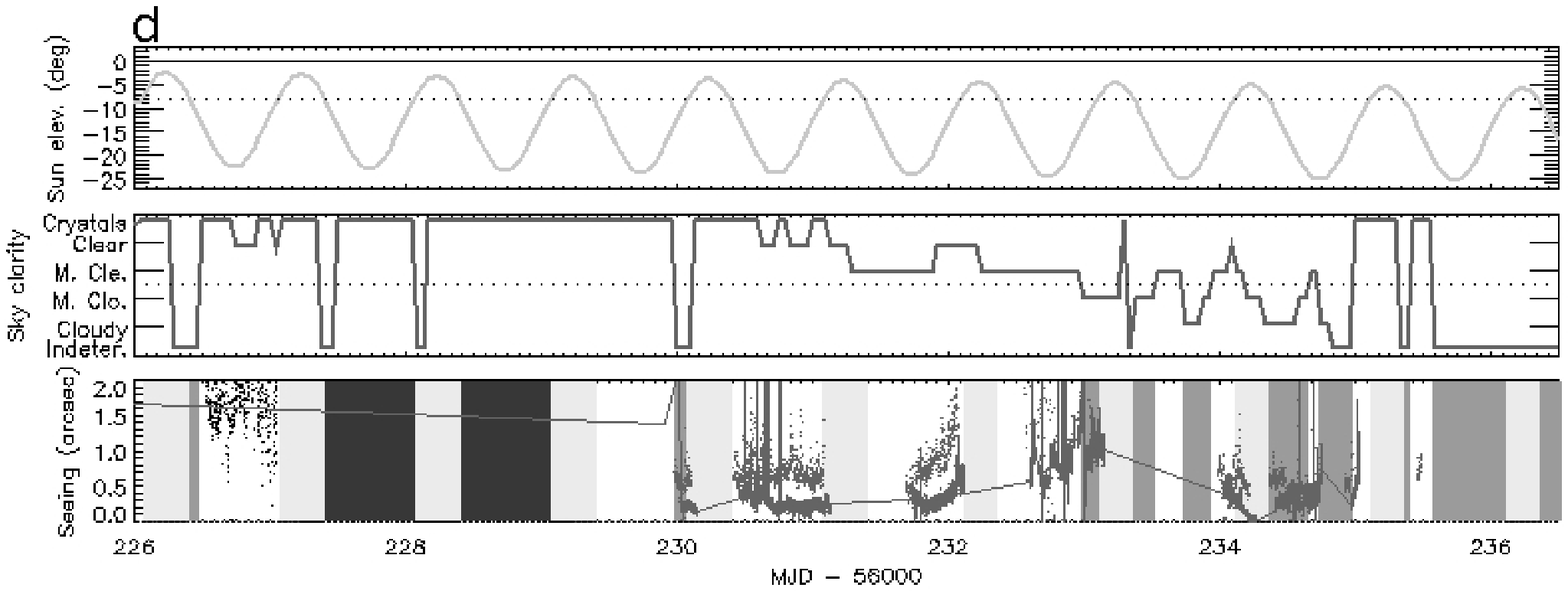}
\caption[]{Continued.}
\end{figure}

\setcounter{figure}{9}

\newpage

\begin{figure}
\plotone{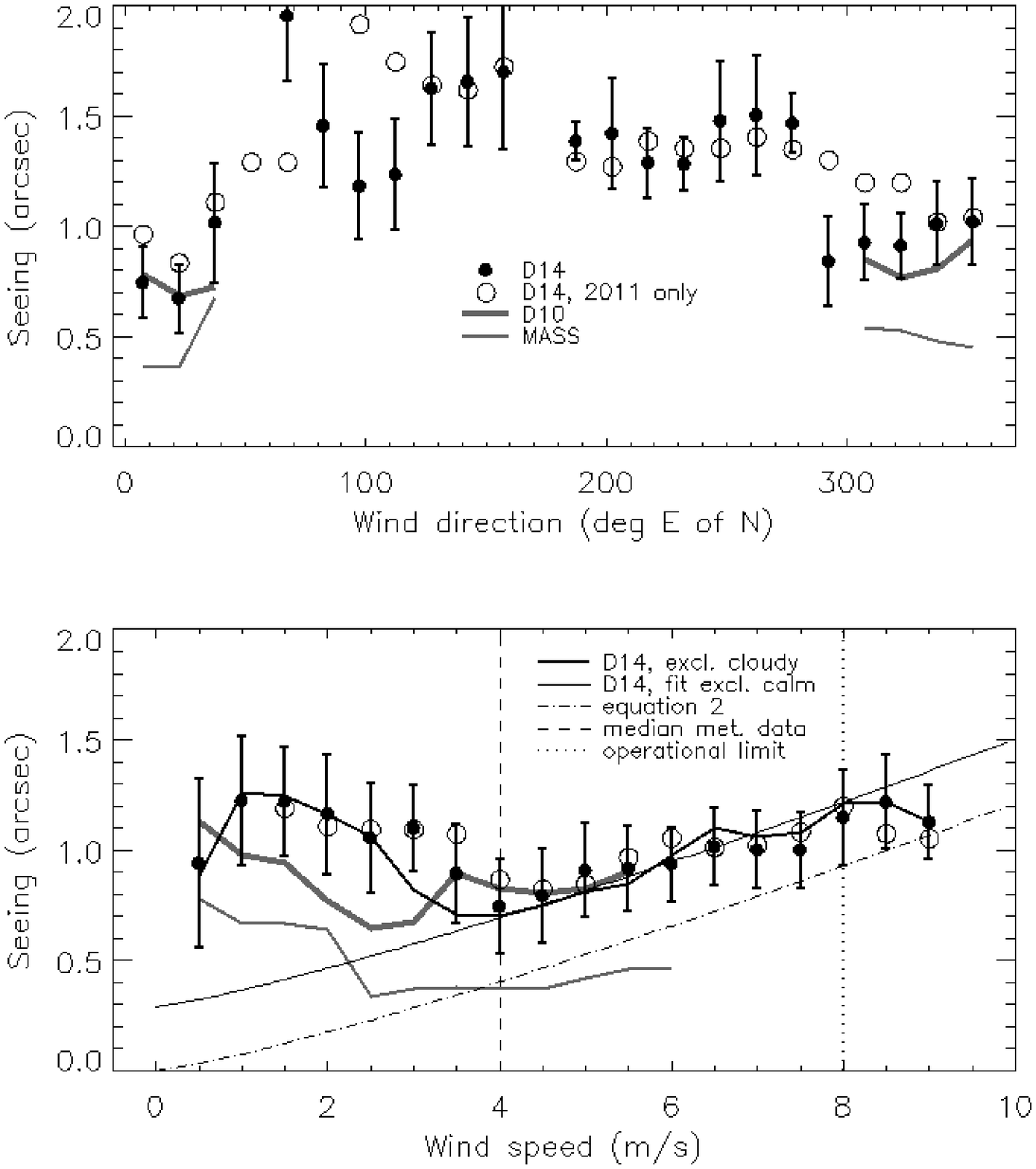}
\caption{All D14 and MASS/D10 results plotted as a function of local ground wind direction and speed; medians in 15\arcdeg~ or $0.5~{\rm m}~{\rm s}^{-1}$ bins; D14 error bars are 1-$\sigma$ limits. Uncertainties are similar for D14 data taken during 2011 only (open circles) and D10 (thick grey lines) and MASS (thin grey lines). DIMM seeing is usually poorer for winds away from north and stronger than $4~{\rm m}~{\rm s}^{-1}$ - or when calm. The situation with MASS may be similar, but without a worsening trend beyond $4~{\rm m}~{\rm s}^{-1}$. A fit of equation 2 to those D14 data above the median wind speed is shown, extrapolated to an intercept at $0~{\rm m}~{\rm s}^{-1}$.}
\label{figure_seeing_correlations}
\end{figure}

\newpage

\begin{figure}
\plotonenarrow{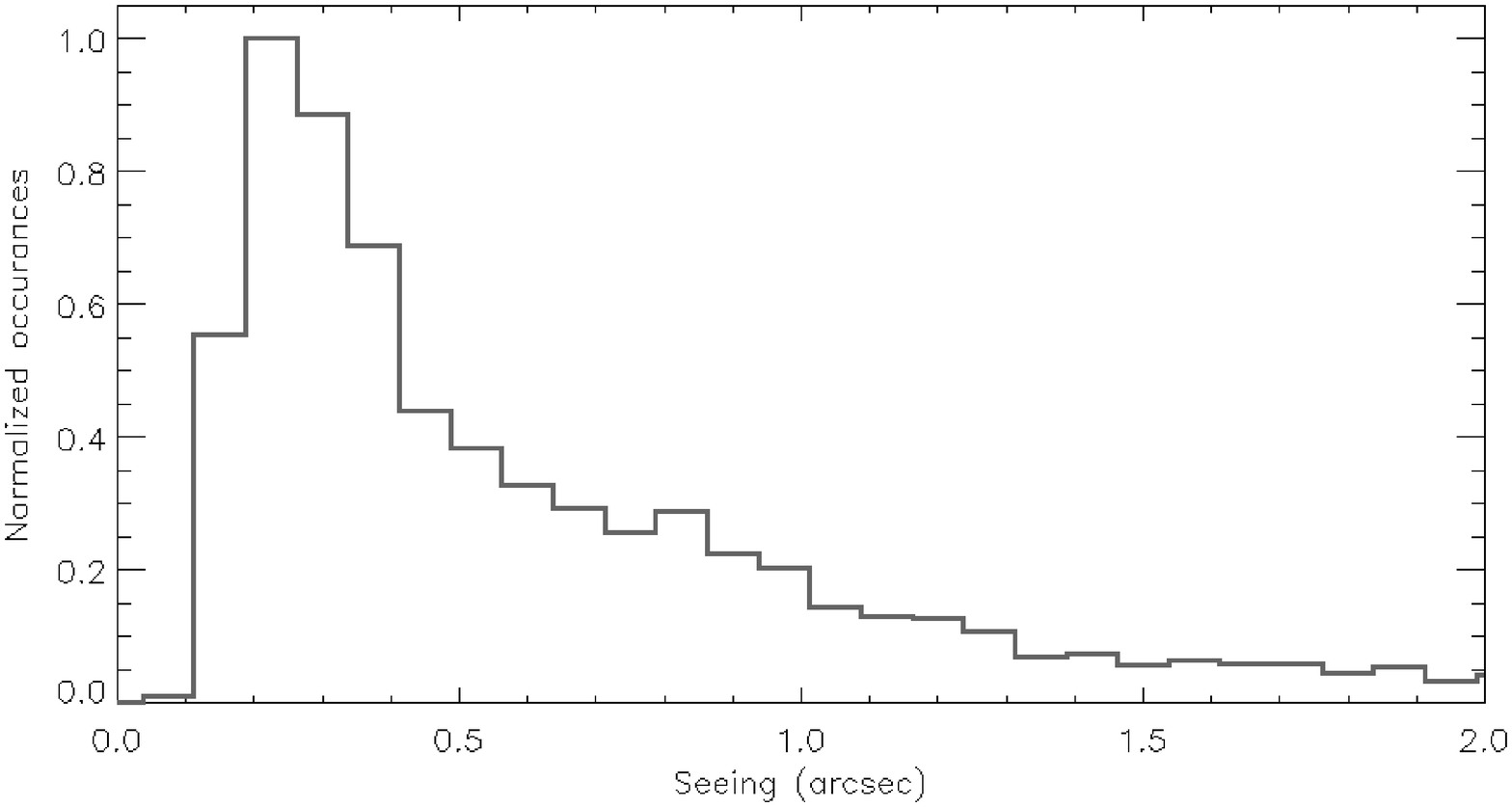}\\
\plotonenarrow{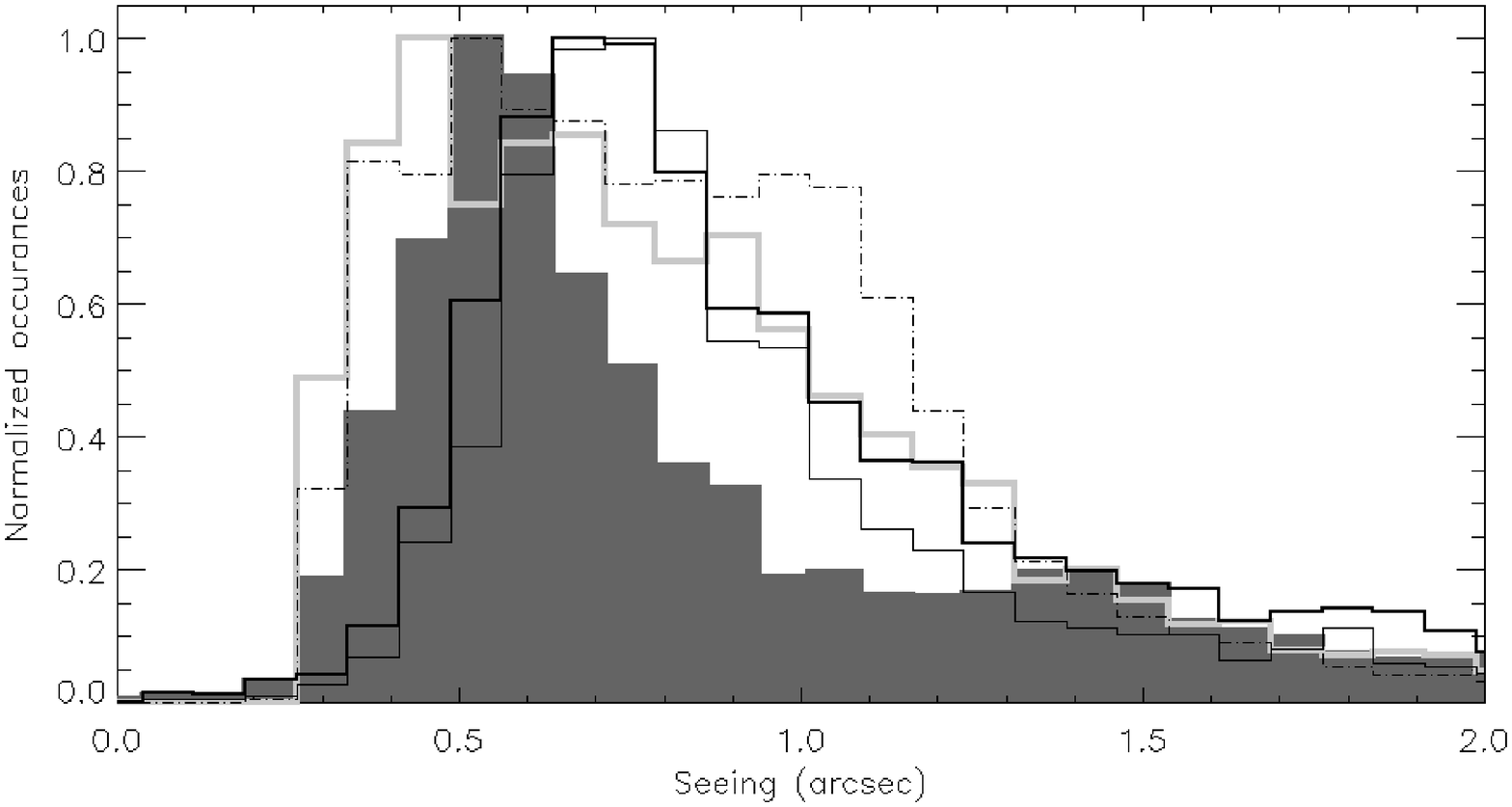}\\
\plotonenarrow{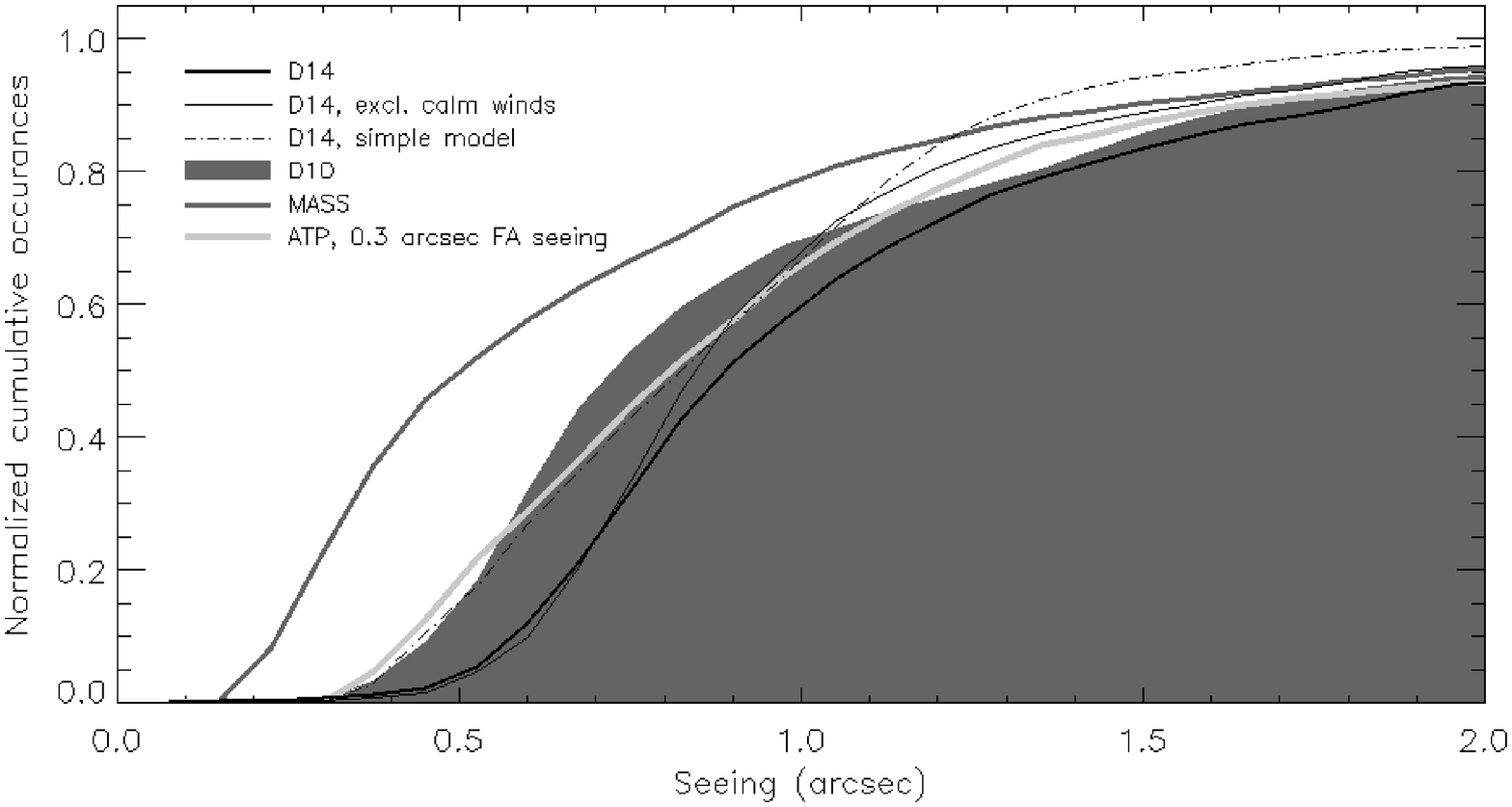}
\caption{Distributions of seeing (top: MASS only, middle: D14/D10/ATP) and cumulative histogram (bottom). D10 measurements are comparable to previous ATP results assuming 0\farcs30 FA seeing \citep{Hickson2013}, with a median similar to D14 data taken in winds under $4~{\rm m}~{\rm s}^{-1}$ as well as a simple model of seeing versus wind speed. See text for details.}
\label{figure_seeing_distributions_model}
\end{figure}

\end{document}